\documentclass[preprint]{aastex}
\usepackage{apjfonts}
\usepackage{graphicx}
\usepackage{epsfig}
\usepackage{float}
\usepackage{amsmath}
\usepackage{natbib}
\usepackage{float}
\usepackage{amssymb}

\usepackage[usenames]{color}

\shorttitle{}
\shortauthors{}

\begin{document}
\title{Diagnosing the burst influence upon accretion in the clocked burster GS 1826--238}

\author{
Long Ji\altaffilmark{1,5}, Shu Zhang\altaffilmark{1}, YuPeng
Chen\altaffilmark{1}, Shuang-Nan Zhang\altaffilmark{1}, Diego F. Torres\altaffilmark{2,3}, Peter
Kretschmar\altaffilmark{4}, Erik Kuulkers\altaffilmark{4}, Jian Li\altaffilmark{3}, Zhi Chang\altaffilmark{1,5}}
\altaffiltext{1}{Laboratory for Particle Astrophysics, Institute of High Energy
Physics, Beijing 100049, China}
\altaffiltext{2}{Instituci\'o Catalana de Recerca i Estudis Avan\c cats (ICREA),
08010 Barcelona, Spain}
\altaffiltext{3}{Institute of Space Sciences (CSIC--IEEC), Campus UAB, Carrer de Can Magrans, s/n 08193, Barcelona, Spain}
\altaffiltext{4}{European Space Astronomy Centre (ESA/ESAC), Science Operations
Department, Villanueva de la Ca\~nada (Madrid), Spain}
\altaffiltext{5}{University of Chinese Academy of Sciences, Beijing 100049, China}


\begin{abstract}
Type-I X-ray bursts on the surface of a neutron star are a unique probe to the accretion in X-ray binary systems. However, we know little about the feedback of the burst emission upon accretion. Hard X-ray shortages and enhancements of the persistent emission at soft X-rays have been observed. To put these findings in context with the aim of understanding the possible mechanism underneath, we investigated 68 bursts seen by {\it RXTE} from the clocked burster GS 1826--238. We diagnosed jointly the burst influence at both soft and hard X-rays, and found that the observations can be described as the {\it CompTT} model with variable normalization, electron temperature and optical depth.
Putting these results in a scenario of coronal Compton cooling via the burst emission would lead to a shortage of the cooling power, which may suggest that additional consideration like the influence of the burst on the corona formation should be accounted for as well.
\end{abstract}

\section{Introduction}
A low-mass X-ray binary (LMXB) consists of a compact star (i.e., a black hole or a neutron star) and a donor.
The donor usually fills its Roche lobe, and thus transfers mass to the compact star via releasing the gravitational potential energy as persistent X-rays \citep[e.g., ][]{Lewin2006}.
Its spectrum can be described as a combination of a thermal and a non-thermal (power-law like) components.
The thermal component is considered as the emission from an optically thick accretion disk, and the power-law component is believed to originate from hot plasmas above or inside the accretion disk \citep{White1988, Mitsuda1989}, which is speculated as being the "corona" \citep[see e.g.][]{Frank2002, Done2007, Zhang2013}.
In case of having a neutron star (NS) as the compact object, the mass accumulated on the NS surface may end up with a violent thermonuclear explosion, in which the integrated flux could reach up to typically $\sim$ $\rm {10}^{39}$--${10}^{40}$\,ergs.
These intense X-ray flares are called type-I X-ray bursts \citep[see e.g.][]{Lewin1993, Galloway2008}.
Being a short intense shower of soft X-rays, type-I X-ray bursts can be used as a probe of the LMXBs (low mass X-ray binaries) accretion via investigating the evolution of the persistent emission along with the runaway of the thermonuclear explosion.
The bursts are expected to have a blackbody spectrum with slight deviations that depend on the burst luminosity \citep[for details, see, e.g., ][]{Suleimanov2011a, Suleimanov2012}.
To extract the burst spectrum, one usually assumes that the persistent emission is unchanged during the burst and one subtracts it off as part of the background \citep[see e.g.][]{Kuulkers2003}.
\citet{Worpel2013}, however, proposed that the persistent emission intensity is variable; thus involving a multiplicative factor $f_{\rm a}$ to the persistent emission for fitting the spectrum during the burst (assuming an unchanged spectral shape, which we shall refer hereafter as the varying persistent flux model).
They found that in photospheric radius expansion bursts (PRE bursts) the best-fit value of the $f_{\rm a}$ is significantly greater than 1, while it seems that the $f_{\rm a}$ values are $\sim$ 1 when the persistent emission is near the Eddington limit.
\citet{intZand2013} found spectral excesses from a blackbody spectrum at both low and high energies in SAX J1808.4--3658, which can be eliminated by using the varying persistent flux model. The value of $f_{\rm a}$ seems to be proportional to the intensity of bursts, while this correlation may reverse at higher burst luminosity in the hard state of LMXBs \citep{Ji2014b}.
In addition, enhanced persistent emission was also reported during a superburst from 4U 1636--536 \citep{Keek2014}.
This shows the significant influence of the burst emission upon the accretion at soft X-rays ($\sim$ 2--10\,keV), although the underlying mechanism is still not understood.
Possible explanations may be Poynting-Robertson effect \citep{Walker1992} or some other reprocessing of burst photons by the accretion disk and the corona \citep{intZand2013}.
In the hard X-rays (e.g., above 30\,keV), on the contrary, the persistent emission during the burst is found to have a clear shortage,
indicating cooling of a corona \citep[see, e.g.,][]{Maccarone2003,Chen2012, Ji2014a} and hence dramatic changes of the persistent spectrum, since the burst emission dominates below 30\,keV.
This issue was partially addressed in previous work \citep{Keek2014}.
They performed a detailed time-resolved spectral analysis for this rare superburst event and found evidences in the decay phase of an enhancement of the overall persistent flux and a reduction of the corona temperature, which could be consistent with a hard X-ray shortage if the persistent spectrum continues up to tens of keV.
Since the superburst event is relatively rare, we explore this issue in more details in this paper by analysing the atoll source NS LMXB GS 1826--238, which shows more frequent X-ray bursts.

GS 1826--238 is a persistent atoll source, which does not exhibit significant spectral evolution of its persistent emission until recently \citep{Nakahira2014,Galloway2008}.
Assuming a distance of 6 kpc, this source always stays in the so-called low hard state with a luminosity of 5$\%$--9$\%$ $L_{\rm edd}$ \citep{Galloway2008}.
Its X-ray emission in a wide energy range of 3--200\,keV can be well described by a pure thermal Comptonization model with a characteristic hot electron temperature, $T_{\rm e}$ $\sim$ 20\,keV \citep{Cocchi2010}.
GS 1826--238 is the famous "clocked burster" because of the recurrence of the bursts being relatively stable over the span of years \citep{Ubertini1999, Galloway2004}.
Also the burst shapes remain quite constant, and the burst peak fluxes have a rather narrow distribution (see, e.g., Table 1).
The mean and standard deviation of the burst peak fluxes are 24.6 and 2.8, respectively.
These properties are quite helpful for reducing the possible systematic errors when one stacks bursts together.
The most distinguishing feature for GS 1826--238 is that it is one of a few atoll XRBs with a hard X-ray shortage being significantly detected during bursts \citep{Ji2014a}.
Since for GS 1826--238 the thermal emission from either accretion disk or NS surface contributes little at energies above 30 keV, such a hard X-ray shortage provides the most clean probe to the change of the non-thermal emission, which likely helps to restrict the spectral model at soft X-rays with respect to the pre-burst emission.

In this paper we derive the evolution of the persistent emission along the burst at soft X-rays, and then diagnose the burst influence on accretion in a broad energy band, using the knowledge of the previously detected hard X-ray shortage.

\section{Observations and data analysis}

We analyzed all {\it Rossi X-ray Timing Explorer (RXTE)} {\it Proportional Counter Array} (PCA) pointing observations and searched for bursts.\footnote{To search for the bursts, we produced 1-s light curves, then select the observations in which the maximum count rate is 500 cts/s larger than the average, and finally checked them manually.}
The bursts exhibiting data losses were excluded.
We also ruled out several bursts for which the Standard 2 mode data were not available.
43 bursts were  previously chosen for studying the hard X-ray shortage in the type I bursts of this source \citep{Ji2014a}.
Since in this analysis the selection criteria are a bit looser, i.e., bursts with incomplete persistent emission (150\,s before and 250\,s after the burst peak) are also considered, 25 more bursts come into the overall sample.
All of the 68 bursts selected for our analysis are shown in Table 1.

In this paper we analyzed the persistent emission using Standard 2 mode data
, while the Event mode data were used for studying the properties of bursts because of their high time resolution.
The software environment was {\it HEAsoft ver 6.15}, including the commands {\it "saextrct"} and {\it "seextract"} to extract light curves and spectra and {\it "pcabackest"} to estimate the background.
The spectral response files were generated by the latest {\it pcarsp V11.7.1}, in which the drifting PCU gain values over the {\it RXTE} mission lifetime have been taken into account.
We have verified that the results are not sensitive to selection of the RXTE epoch.
The spectral analysis was performed by using {\it Xspec 12.8} and its Python wrapper {\it PyXspec}.
During spectral fits, the used energy band was 3--20\,keV, above which the effective area of {\it PCU} drops off rapidly.
To avoid the possible bias in estimation of parameters because of the non-Gaussian distribution in some channels with small numbers of counts, a Churazov weight\footnote{see https://asd.gsfc.nasa.gov/XSPECwiki/low\_count\_spectra} was used in the spectral analysis.
We performed time-resolved analysis on each burst, by adopting the same spectral fitting procedure as described in \citet{Ji2014b}.
In brief, we performed time-resolved analysis for the individual bursts around their peaks.
To improve the statistics, we selected variable time intervals so that the total counts in each time bin exceeds
 5000 photons.
As a result, the bin size has to gradually increase toward the burst tail because of the lower count rate.
The varying persistent flux model, i.e., $wabs*(bbodyrad+f_a \times persistent\ flux$) was employed to fit the spectra in the 3--20\,keV band, in which the $bbodyrad$ represented the intrinsic burst component, while $f_{\rm a} \times persistent\ flux$ represented the variable persistent emission \citep[see, e.g.,][for details]{Worpel2013}.
The "wabs" component was used to describe the interstellar absorption; the hydrogen column density was fixed to 0.3 $\times$ ${10}^{22}$ atoms\ $\rm {cm}^{-2} $ \citep{Cocchi2010}.
Our results are independent of the photoelectric absorption model due to the fact that its major influence happen at energies below 3\,keV.
The varying persistent flux model was chosen because the short burst exposure does not allow measuring all parameters of the relatively weak persistent emission (see Table~1).
Here the $persistent\ flux$ was estimated by fitting the spectrum before each burst with a phenomenological model, such as $wabs*(powerlaw+gauss)$ or $wabs*(bbodyrad+powerlaw+gauss)$, where the Gaussian component was always set at 6.4\,keV with an arbitrary width of 0.5\,keV.
More physical models, e.g., {\it CompTT}, would lead to large parameter errors due to {\it RXTE/PCA}'s limited bandpass, but was addressed in the subsequent simulations.
An average reduced-${\chi}^2$ of $\sim$ 1.3 (45 dof) shows that it is appropriate to describe phenomenologically the persistent emission with the adopted model.
%

\section{Results}

\subsection{$f_{\rm a}$ evolution and correlation with burst flux}

As mentioned above, we performed time-resolved analysis for individual bursts to obtain the curves of the burst flux and $f_{\rm a}$ factor.
For individual bursts, however, the possible change of the $f_{\rm a}$ parameter is not very significant due to relatively large error bars (see Figure 1 in \citet{Ji2014b}).
In order to improve the statistics, we investigated the average evolution of $f_{\rm a}$ by summing up all the bursts.
In practice, we took the peak flux of each burst as a time reference (t=0\,s) and averaged over bins of 3 seconds.
Figure~\ref{multiburst} shows that at the rising phase of bursts, the $f_{\rm a}$ value rapidly increases (O to A).
Then, along with a further increasing of the average burst flux, the $f_{\rm a}$ value decreases and reaches a value of around 1 at time of the burst flux peak (noted as point B in Figure~\ref{multiburst}).
Finally, during the averaged burst decay, $f_{\rm a}$ gradually returns to a value (point C) comparable to that at point A, and then smoothly decreases.
We note that if we divide the bursts into two groups along time and then calculate $f_{\rm a}$ curves individually to check their consistency, the ${\chi}^2$ test leads to a reduced-${\chi}^2$ 51.83(46 dof), which indicates that the slight changes of the shape of bursts over several years \citep{Galloway2004,Galloway2012} have little influence on our results.
For comparison, Fig.~\ref{multiburst} also presents the averaged hard X-ray (30-50 keV) lightcurve during bursts observed by {\it RXTE/PCA} using the Event mode data \citep{Ji2014a}.
\footnote{We note that compared with the hard X-ray lightcurve, 25 more bursts are used to produce the soft X-ray and the $f_{\rm a}$ lightcurves with smaller error bars (i.e., upper panel in Figure 1). The soft X-ray and the $f_{\rm a}$ lightcurves are changed little if we use identical selection of bursts.}
The hard X-ray shortage is significant around the burst peak and gradually recovers to its initial level.

To illustrate the $f_{\rm a}$ evolution against flux, we present these data also in the Figure~\ref{multiburst2}.
The correlation is significantly different between the cases when the burst flux is larger and smaller than $\sim {10}^{-8}$\,erg\ $\rm s^{-1}\ {cm}^{-2}$.
The turnover is smooth instead of a sharp peak.
The Pearson correlation coefficient is 0.87 with a p-value of $2.4 \times {10}^{-15}$ for the former, but -0.95 with a p-value of $2.6 \times {10}^{-7}$ for the latter.
This result strongly confirms what was reported previously: $f_{\rm a}$ is highly dependent on the burst luminosity \citep{Worpel2013,intZand2013,Keek2014}.
Although a quite similar $f_{\rm a}$ trend was reported previously in the hard state of 4U 1608--522 \citep{Ji2014b},
the fact that no significant shortage of the persistent hard X-ray emission is observed in this source \citep{Ji2014c}, prevents us from a joint diagnosis of the $f_{\rm a}$ evolution over a broad energy band.

As shown in \citet{Ji2014a} for GS 1826-238, the hard X-ray shortages for the persistent emission are still significant around points A and C of the bursts, suggesting a cooled corona.
However, at that time, the $f_{\rm a}$ curve at soft X-rays show large enhancements of the persistent emission ($f_{\rm a}$ $\sim$ 1.8).
Around the burst peak, both the hard X-ray flux and the $f_{\rm a}$ present a similar trend of anti-correlation with the burst flux, which likely suggests a common cause.
An enhancement in $f_{\rm a}$ can be the result of, e.g., the intrinsic spectrum that deviates from a blackbody slightly \citep{Suleimanov2011a,Suleimanov2012} or the reflection components \citep{Ballantyne2004,Ballantyne2005,Keek2014b}, which might lead to a biased estimate of $f_{\rm a}$ when the "bbodyrad" model is used.

In what follows we diagnose these possible mechanisms in the light of the possible burst influence on the persistent emission, using the results of $f_{\rm a}$ as derived at soft X-rays in this paper and of the persistent emission shortage reported previously at hard X-rays \citep{Ji2014a}.

\subsection{Diagnosing the burst influence}
Here we adopt the "{\it burstatmo}" model to describe the intrinsic spectra of bursts instead of the simple {\it bbodyrad} \citep[see,][]{Suleimanov2011a,Suleimanov2012}.
The "burstatmo" model represents the atmospheres of neutron stars and emergent spectra in a plane-parallel, hydrostatic, and local thermodynamic equilibrium (LTE) approximation.
Hence, we performed again the above-described analysis, but now using a slightly different model: $wabs*(burstatmo+f_{\rm a} \times persistent\ flux$).
There are three parameters in the {\it burstatmo} model, i.e., the radius and luminosity of the neutron star, and a normalization.
We can not simultaneously constrain these parameters due to the large error bars and chose to fix the less impacting parameter--the radius of the neutron star--at 10\,km.
If the increasing value of $f_{\rm a}$ up to $\sim$ 1.8 originates from the deviation of the intrinsic burst spectrum from a blackbody, after using "burstatmo" model, $f_{\rm a}$ should be around 1.
However, we found that the resulting $f_{\rm a}$ is very similar to the case when {\it bbodyrad} model was used (i.e., similar to the Figure ~\ref{multiburst} but with slightly bigger error bars).
In addition, the "burstatmo" model does not improve the goodness of fit.
The reduced ${\chi}^2$ is 1.10 (18 dof), which is comparable with 1.08/18(dof)using the model
$wabs*(bbodyrad+f_{\rm a} \times persistent\ flux$).
In order to further rule out the possibility that the $f_{\rm a} > 1$ originates from the bias of fitting due to the spectral subtle deviations between the intrinsic spectra (i.e., {\it burstatmo}) and phenomenological {\it bbodyrad} model, we used the {\it burstatmo} model to produce faked spectra with assumed burst luminosities 0.5 and 1, respectively, and then fitted them by varying persistent flux model using {\it bbodyrad} model.
The resulting best-fit values of $f_{\rm a}$ are 1.31 and $-4.01$, respectively, which are inconsistent with the observed values (i.e., 1.6 and 1.1 with error bars of $\sim$ 0.1) shown in Figure~\ref{multiburst}.

Alternatively, the reflection during bursts produced by the accretion disk represents viable possibilities for influencing the persistent flux \citep{Ballantyne2004, Ballantyne2005, Keek2014b}.
These were detected during superbursts \citep{intZand2010, Keek2014} or hinted at in normal bursts \citep{Galloway2008}.
For individual bursts, the statistics are not good enough to search for possible reflection components.
Therefore, we stacked the residuals derived from the spectral analysis using the varying persistent flux model.
If the reflection component does exist, one would expect some structure in the residuals.
The analyses were performed at three instances (t=0\,s, 20\,s, 60\,s) with a time bin of $\sim$ 1\,s for the different phases of the (average) burst.
We found, however, no additional structures suggestive of emission lines or edges.
The excess around 6.4\,keV is hardly visible with a significance level below 2\,$\sigma$.

The hard X-ray shortage during bursts in GS 1826--238 (and other LMXBs) is indicative of cooling of the corona \citep{Maccarone2003,Ji2014a}.
The result reported from a superburst in 4U 1636--536 also showed evidence in the burst tail for changes of the electron temperature $T_{\rm e}$ in the corona when the blackbody normalization was fixed \citep{Keek2014}.
We investigated whether the $f_{\rm a}$ variability shown in Figure~\ref{multiburst} corresponds to any corona cooling by
performing simulations.
We set free the electron temperature $T_{\rm e}$ and produced some faked spectra, and fitted them using the varying persistent flux model to look into a possible influence on $f_{\rm a}$ factor.
We used the  model {\it CompTT} to describe the persistent flux; the model parameters (i.e., $T_{\rm 0}$=1.02, $T_{\rm e}$=21.5\,keV and $\tau$=3.81) were adopted from the report of \citet{Cocchi2010}.
We note that a slightly different model, e.g., {\it cutoffpl} or {\it comptb}, barely affects our analysis.
Assuming the spectrum during the bursts is $wabs*(bbodyrad+CompTT)$ with variable $T_{\rm e}$, we produced faked spectra by using the {\it Xspec} command {\it fakeit}, where the {\it bbodyrad} components are set at typical values T=2\,keV and $norm$=100.
Then we used the varying persistent flux model to fit these faked spectra, i.e., assuming the spectrum remains constant in shape but varies in normalization.
The resulting value $f_{\rm a}$, as illustrated in Figure~\ref{shortageWithTe}, clearly indicates that a lower temperature $T_{\rm e}$ tends to result in a smaller $f_{\rm a}$.

Considering that the 3 additional independent variables in the coronal model {\it CompTT}, i.e., the normalization ($N$), the optical depth ($\tau$), and the temperature of the seed photons ($T_{\rm 0}$), are also expected to vary under the shower of soft X-rays from bursts, we further investigated if the $f_{\rm a}$ evolution can be connected to  variations in these parameters.
For simplicity, we first kept $\tau$ and $T_{\rm 0}$ unchanged during bursts, but set free the variables $N$ and $T_{\rm e}$; see below for a discussion of variations in the other parameters.
We performed simulations by producing faked spectra based on the model $wabs*(bbodyrad+N \times CompTT(T_{\rm e}))$, in which there are two free variables: $N$ and $T_{\rm e}$.
We again used the varying persistent flux model to fit these faked spectra, like done before.
The normalization of {\it CompTT} is fixed to $\rm 4.78 \times {10}^{-3}$, for which the inferred counts rate is consistent with the hard X-ray (30--50\,keV) observations, i.e., 1.69\,$\rm cts\ s^{-1}\ {pcu}^{-1}$.
The assumed temperature and normalization in the {\it bbodyrad} model are 2\,keV and 100, respectively, which corresponds to the values around the burst peaks; note that the simulations are insensitive to the assumed blackbody parameters.
The result is illustrated in the left panel of Figure~\ref{contour}.
The colorbar in that figure shows the value of $f_{\rm a}$, whereas the black lines show the contours where $f_{\rm a}$ equals to 0, 1 and 2, respectively.
We note that in Figure~\ref{contour} the $f_{\rm a}$ is just a result of using varying persistent flux model, and does not reveal physical meanings, i.e., the enhanced (suppressed) accretion rate when the $f_{\rm a}$ is larger (smaller) than 1.
The result suggests that larger $N$ and $T_{\rm e}$ lead to a larger $f_{\rm a}$ value.
This is not surprising since both of the $N$ and $f_{\rm a}$ are a normalization; and $T_{\rm e}$ will affect the efficiency of the up-Compton scattering.
Thus, the scenario of corona cooling via bursts can be sketched as follows.
At the beginning of the burst, the normalization (x-axis, $N$) increases rapidly, which phenomenologically accounts for the $f_{\rm a}$ > 1 (O-->A, point O represents the initial values before the bursts, i.e., $T_{\rm e}=21.5\,keV$ and $N$=1).
This process may be accompanied by a moderate decrease of the $T_{\rm e}$.
Then, $T_{\rm e}$ drops sharply and dominates the burst influence by largely reducing $f_{\rm a}$ (A-->B and B-->C).
In the cooling tails the process might be a rough inverse of the O-->A process, but with a dominance of corona heating over than the cooling.\footnote{Note that Figure~\ref{contour} is just a schematic diagram and the positions of the points (A/C, B) are not precise; they only representing the values of $f_{\rm a}$ $\sim$ 1.8, and $\sim$ 1, respectively.}
We note that the burst flux and the $f_{\rm a}$ value at point A are comparable with these at point C, while its hard X-rays seems to be slightly stronger, which may imply a little larger $T_{\rm e}$ at point A.

As mentioned above, the hard X-ray shortage in the bursts of GS 1826--238 is observed to be rather robust \citep{Ji2014a}.
This provides a unique opportunity for testing the cooling scenario speculated above, via checking if the observed hard X-ray shortage is consistent with the persistent spectral model derived from the varying persistent flux model.
For a given persistent model and assumed parameters, one can deduce the hard X-ray count rates by convolving it with {\it PCA}'s response matrix, and then compare to the results using observations directly.
Therefore, the observed hard X-ray shortage can impose a strong constraint in the parameter space of the persistent spectral model.
The hard X-ray count rate at 30--50 keV before the bursts is $\sim$ 1.69 $\pm$ 0.1\,$\rm cts\ s^{-1}\ {pcu}^{-1}$, and is reduced to a minimum $\sim$ 0.8 $\pm$ 0.1\,$\rm cts\ s^{-1}\ {pcu}^{-1}$ around the burst peak.
Assuming the coronal spectrum is $N \times CompTT(T_{\rm e})$, these hard X-ray shortages correspond to the upper and lower cross hatches in the left panel of Figure~\ref{contour} respectively, between which is the constrained parameter space of $N$ and $T_{\rm e}$.
We found that for points A and C the parameter spaces having hard X-rays between $\sim$ 0.8 $\pm$ 0.1\,$\rm cts\ s^{-1}\ {pcu}^{-1}$ and $\sim$ 1.69 $\pm$ 0.1\,$\rm cts\ s^{-1}\ {pcu}^{-1}$ (i.e., between the two hatches) do not overlap the region in which the $f_{\rm a}$ is $\sim$ 1.8.
This implies that the {\it CompTT} model with variable $N$ and $T_{\rm e}$ can not reconcile the observations at both hard and soft X-rays.

To solve such an incompatibility, one may consider additionally the other two variables in {\it CompTT}, i.e., the temperature of seed photons $T_{\rm 0}$ and the optical depth $\tau$.
We found that the temperature of seed photons influences little both the hard X-ray shortage and the $f_{\rm a}$ evolution, probably because $T_{\rm 0}$ (which is $\sim$ 1\,keV) is comparable to the burst temperature ( $\sim$ 1--2\,keV).
$\tau$ , however, has a great effect on the persistent spectrum, especially at the hard X-rays.
When $\tau$ decreases from an initial value of 3.81 \citep{Cocchi2010} to $\sim$ 2, the parameter region inferred from the hard X-rays shortage (at 30--50\,keV) is compatible with that estimated from $f_{\rm a}$ evolution (at 3--20\,keV).
This result is shown in the right panel of Figure~\ref{contour}, and hints at a variable optical depth $\tau$ as a consequence of the burst.

\section{Discussion and Summary}

Feedback of type-I X-ray bursts onto the accretion material surrounding the compact star in LMXBs has been predicted earlier \citep{Walker1989, Walker1992, Ballantyne2004, Ballantyne2005, Inogamov1999, Inogamov2010, Kluzniak2013}.
However, due to the generally short exposure of bursts, it is hard to test them in individual events.
Using {\it RXTE} observations of GS 1826-238 spanning seventeen years, we were able to study in detail the overall changes of the persistent emission while bursting. This study can be done due to its constant burst shape \citep[e.g.,][]{Thompson2005} and dominance of the coronal emission in the hard state, and, therefore, significant non-thermal radiation above $\sim$ 30\,keV \citep[e.g.,][]{Galloway2004}.
Following \citet{Worpel2013}, we employed the varying persistent flux model to fit the spectra in the energy band 3--20 keV during bursts.
The $f_{\rm a}$ is correlated with the burst flux up to $\sim$ $\rm {10}^{-8}\,erg\ s^{-1}\ {cm}^{2}$
(which corresponds to $\sim$ $\rm 0.8 \times {10}^{38}\,erg\ s^{-1}$ assuming a distance of 6 kpc and a gravitational redshift of 1.3),
while above this value it is anti-correlated.
This confirms the previous report using the hard state of 4U 1608--522 \citep{Worpel2013,Ji2014b}.
In 4U 1608-522, however, no hard X-ray shortages were found \citep{Ji2014c}.\footnote{We note that the lack of a hard X-ray shortage might be caused by e.g., the contamination of bursts having high color temperature, or the lack of significant non-thermal emission. For details, see \citet{Ji2014c}.}

We found that, although the varying persistent flux model is successful in fitting the persistent spectrum during bursts, especially in the 3--20\,keV energy band, more physical mechanisms are required to account for the contemporary hard X-ray shortage.
We considered a variety of possibilities among which only the inverse Compton cooling of the corona seems reasonable.
We found that an increase of the coronal model {\it compTT} normalization ($N$) by a factor of 2--3 can account for the $f_{\rm a}$ enhancement with respect to the pre-burst emission.
This increased $N$ might result from the increasing seed photons provided by type-I X-ray bursts, or the Poynting-Robertson drag effect \citep{Walker1992}.
We note that the luminosity of the burst overwhelms that of the pre-burst accretion disk by a factor of more than 10, while the {\it CompTT} normalization only increases by a factor of 2--3.
This suggests that the interaction between the accretion disk and the corona is more efficient, which would be the case if the corona "sandwiches" the disk \citep{White1988}, but not when it is spherical in the inner side of the truncated disk \citep{Mitsuda1989}.
This result is in agreement with the recent report of \citet{Zhang2014}, who found that the spectra of dipping (i.e., high-inclination) LMXBs, regardless of their spectral states, exhibit stronger comptonized emission, suggesting that coronae may have disk-like or oblate shapes.

\begin{figure}
  \centering
  \includegraphics[width=3.2in]{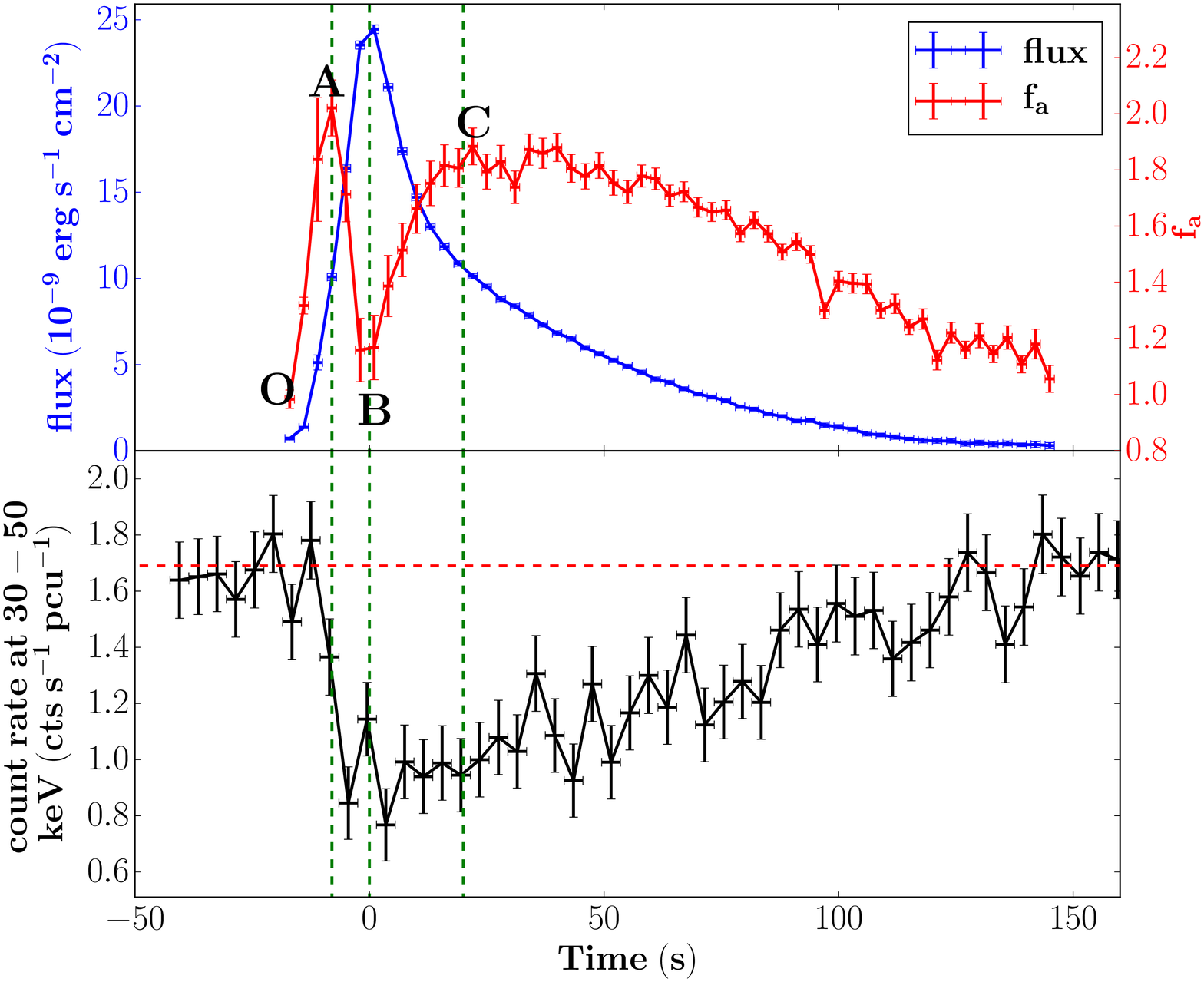}
\caption{The evolution of the averaged flux, value of $f_{\rm a}$ (upper panel) and hard X-rays at 30--50 keV (bottom panel).
 The red horizontal line represents that the count rate at 30--50 keV equals to 1.69 $cts\ s^{-1}\ {pcu}^{-1}$.
 The green vertical lines represent that the time equals to -8, 0, 20 s, respectively.
 }
\label{multiburst}
\end{figure}

\begin{figure}
  \centering
  \includegraphics[width=3.2in]{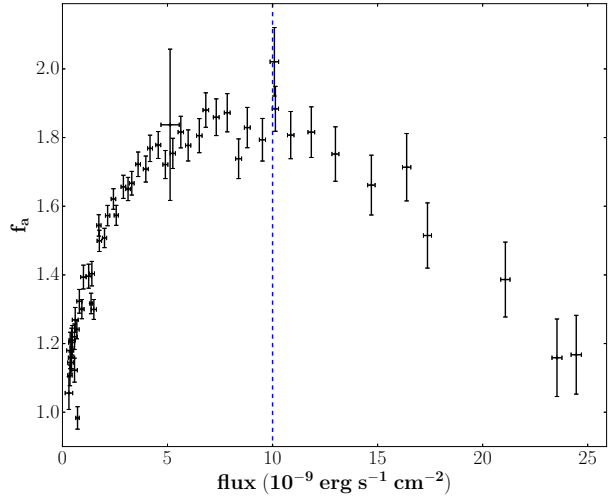}
\caption{The burst flux vs. the value of $f_{\rm a}$.
The burst flux is estimated as $4\pi \sigma T^{4} N$, where $\sigma$ is the Stefan--Boltzmann constant, $T$ and $N$ are the best-fit parameters of blackbody model "{\it bbodyrad}", respectively.
 The dash line represents the flux equals to 0.8 $\rm \times {10}^{38}$\, erg $\rm s^{-1}$.
 }
\label{multiburst2}
\end{figure}

\begin{figure}[!htbp]
  \centering
  \includegraphics[width=3.2in]{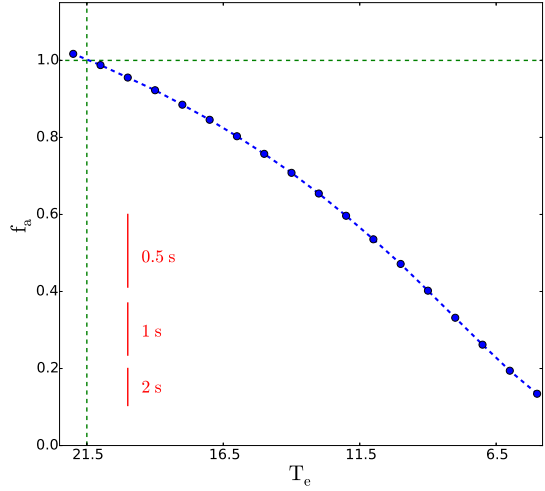}
  \caption{
  The expected $f_{\rm a}$ vs. the electron temperature $T_{\rm e}$.
  The dashed lines show the initial parameters before bursts, i.e., $T_{\rm e}$=21.5 and $f_{\rm a}$=1.
  The red lines represent the characteristic error bars when stacking the burst spectra with exposure time 0.5\,s, 1\,s and 2\,s, respectively.
  }
\label{shortageWithTe}
\end{figure}

\begin{figure}[!htbp]
  \centering
  \includegraphics[width=3.2in]{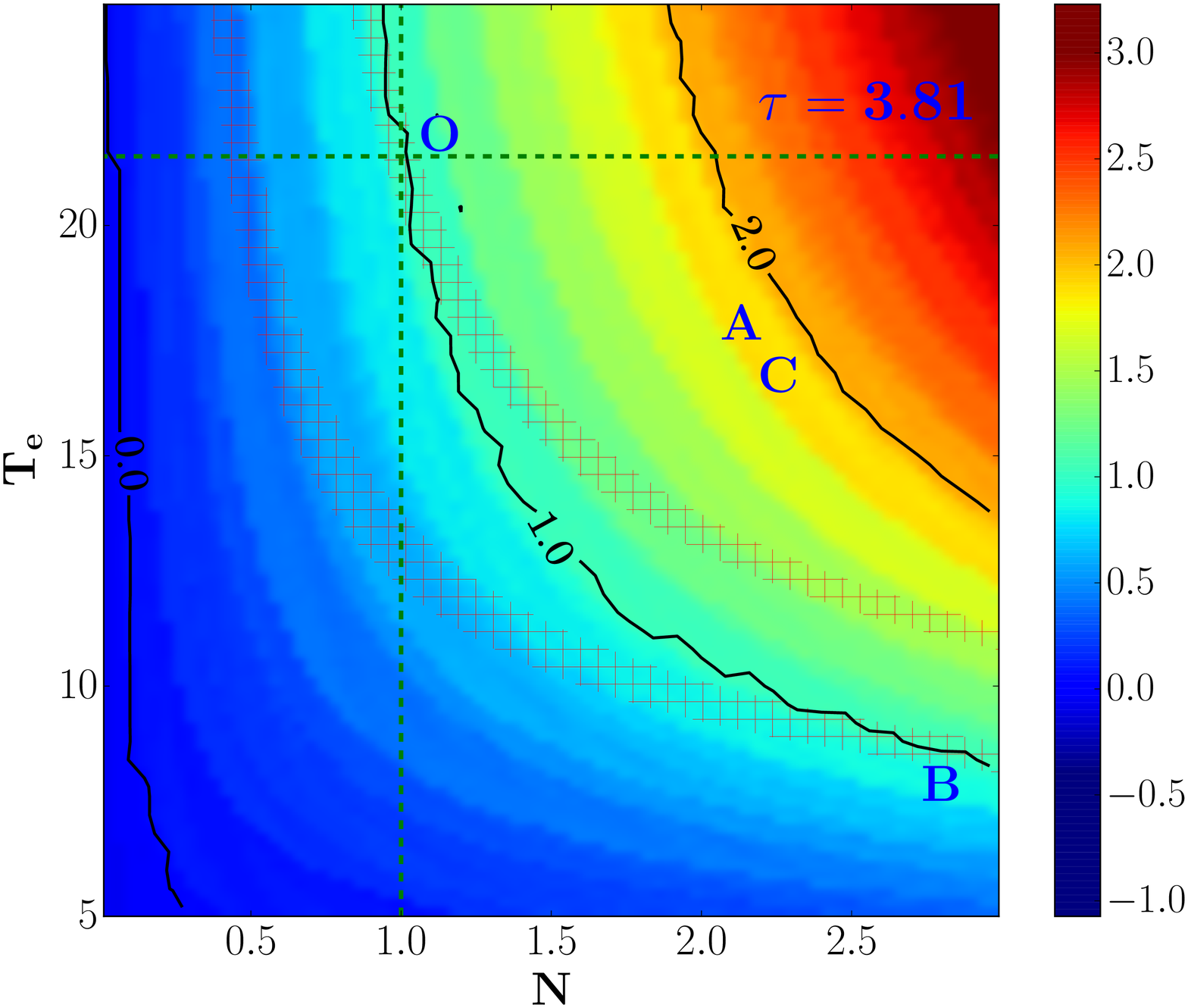}
  \includegraphics[width=3.2in]{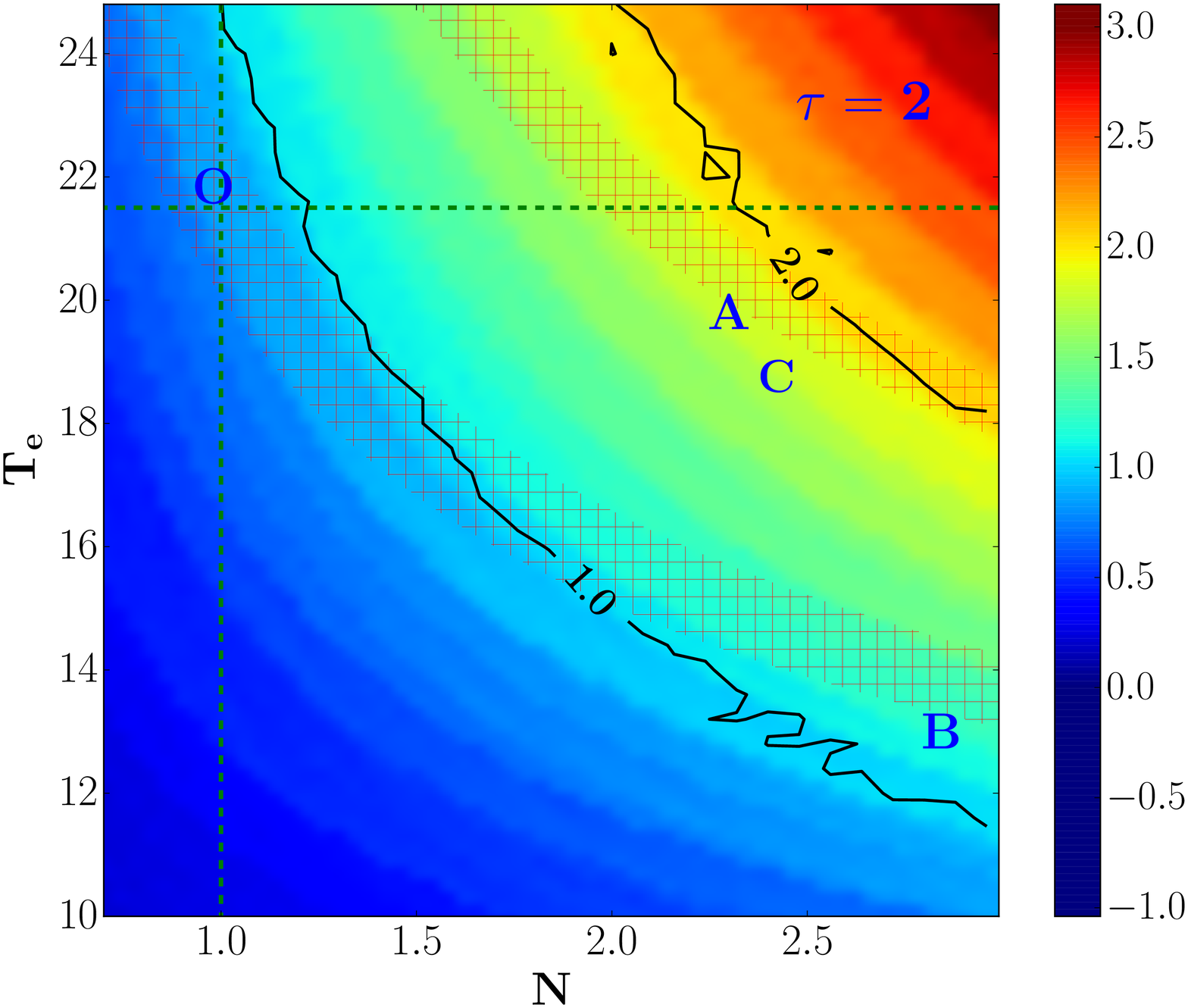}
  \caption{
  The resulting $f_{\rm a}$ by assuming the $\tau$ = 3.81 and 2 for the left and right panel, respectively.
  The colorbar shows the intensity of the $f_{\rm a}$ and the black lines shows the contours where the $f_{\rm a}$ equals to 0, 1 and 2, respectively.
  The two cross hatches enclose the possible region of $N$ and the $T_{\rm e}$, which are deduced from the hard X-rays shortages in Figure~2 in \citet{Ji2014a}.
  The green lines show the initial state before bursts, $f_{\rm a}$ =1 and $T_{\rm e}$=21.5 keV.
  }
\label{contour}
\end{figure}

Around the point B in Figure~\ref{multiburst}, $f_{\rm a}$ has a trend similar to that of the strength of the hard X-ray shortage.
This hints at a sufficiently cooled corona when the burst emission is intense.
Figure~\ref{contour} shows that the electron temperature $T_{\rm e}$ may account for the $f_{\rm a}$ valley observed around the burst peak.
Such a $T_{\rm e}$ reduction/recovery feature was observed as well in the decay phase of the superburst from 4U 1636--536  \citep{Keek2014}, which also points to corona cooling under a shower of soft X-rays.
We find, however, that to reconcile the hard X-ray shortage with the $f_{\rm a}$ evolution at soft X-rays, one has to take into account a smaller optical depth $\tau$.
The Comptonised spectrum can be used to estimate the cooling power, i.e., the power of losing energy of the corona under the shower of burst photons.
Here we estimated the cooling power as the difference between the incident flux and the emergent flux after the Compton process.
The emergent flux is estimated by using {\it Xspec} command {\it "flux"} with the parameters specified below\footnote{The {\it "flux"} command can be used to calculate the flux of the specified model, parameters and energy ranges. For details, see https://heasarc.gsfc.nasa.gov/xanadu/xspec/manual/XSflux.html}, and the incident flux is inferred by analytic estimation, i.e., the emergent energy of photons is $\sim$ $e^{\frac{4kT_{\rm e}}{m_{\rm e} c^2} Max(\tau, {\tau}^2)}$ times larger than the incident energy \citep{Rybicki1979}.
If we assume that around the burst peaks (point B) $T_{\rm e}$ $\sim$ 13\,keV and $\tau \sim$ 2, the inferred cooling power is $6.8 \times {10}^{34} {(\frac{D}{6\,kpc})}^2 ergs\ s^{-1}$ in the energy band 1--100 keV, which is only $\sim$ 8\% of the cooling power before bursts.
The uncertainties of the cooling power is difficult to be constrained.
If we artificially assume the error bars of $T_{\rm e}$ and $\tau$ are $\sim$ 3 keV and 1 at 90\% confidence levels, respectively, and they are independent, the 5\% lower and upper boundaries of the inferred distribution of the cooling power are 1.5\% and 24.0\% of the cooling power before bursts, respectively.
This cooling power turns out to be not sufficient to account for the decreased temperature of electrons in the corona \citep{Keek2014}.
Thus we speculate that a possible solution might be to take into account an additional mechanism, i.e., the corona formation, which may be influenced as well by the burst.
This may be reminiscent of the XRB outbursts in their spectral transition from hard state to soft state, where the corona is eventually diminished along with the increasing accretion rate.
If one assumes that the changes of the electron temperature are the result of the balance between the Compton cooling and heating mechanism, a suppressed heating process might be a solution to the above contradiction.
However, the heating mechanism currently is still poorly known.
In evaporation models, both the disk and corona are individually powered by the release of gravitational energy associated with the accretion of matter affected through viscous stresses, which depends strongly on the viscosity parameter ($\alpha$) \citep{Meyer1994,Liu2013}.
Alternatively, in magnetic reconnection models, the heating power is $\sim$ $\frac{B^{2}}{4\pi}{V}_{A}$, where $V_{A}$ is the Alfv\'{e}n speed \citep{Liu2002}.
Therefore, if at any point in time the system is in approximate equilibrium, i.e., the heating power equals to the cooling power, the viscosity parameter $\alpha$ or the Alfv\'{e}n speed $V_{A}$ is expected to change dramatically during bursts.
We caution that this scenario might be oversimplified.
In outburst transitions, which is also attributed as the cooling and reheating of the corona, the low state to high state spectral transition at the initial rise of outbursts occurs at a luminosity 5 times greater than that of the transition from high state to low state at the declining stage \citep{Maccarone2003b}, which is known as "hysteresis effect" .
This means that the seed photons' flux might be not the only factor that can influence the cooling and reheating of the corona, and some other unknown processes should be taken into account \citep{Begelman2014}.
We speculate that during transitions from the hard state to the soft state in outbursts of LMXBs, the corona is cooled by the increasing soft seed photons emitted from the accretion disk, which in general leads to an increased value of $\tau$ \citep[for details, see, e.g.,][]{Narayan1995,Esin1997}.
This seems inconsistent with what we infer for GS 1826--238.
As shown in the left panel of Figure~\ref{contour}, the main inconsistency that results from $f_{\rm a}$ and the hard X-ray shortage is at the points A and C.
A smaller optical depth of the corona has to be introduced to account for the difference of the persistent emission modifications between the soft and the hard X-rays.
We speculate that this 'discrepancy' might arise from the different optical depths of the corona for the seed photons having different origins.
Considering that if the corona is geometrically thick above the accretion disk, but exists only within a small radial radius, the optical depth for the seed photons from the accretion disk would be larger for these photons from type-I X bursts.
In addition, the seed photons from type-I bursts would lead to significant Poynting-Robertson drag \citep{Walker1992}.
This may imply that during bursts the inner part of the corona vanishes and becomes thinner in the radial direction \citep{Walker1992}, leading to a smaller optical depth.

\acknowledgments
We acknowledge support from the Chinese NSFC 11473027, 11133002, 11103020, XTP project XDA 04060604
and the Strategic Priority Research Program "The Emergence of Cosmological Structures" of the Chinese Academy of Sciences, Grant No. XDB09000000 and also by the Strategic Priority Research Program on Space Science,the Chinese Academy of Sciences, Grant No.XDA04010300.
DFT work is supported by grant AYA2012-39303, and further acknowledges the Chinese Academy of Sciences visiting professorship program 2013-T2J0007.
We thank V. Suleimanov for providing the {\it Xspec} code for the atmosphere model of neutron stars.

\mbox{}

{}

\begin{table*}[t]
\tiny
\caption{The columns denote the observation ID, the time when type-I bursts occurred, persistent flux and unabsorbed peak flux of bursts at 3--20 keV.
The persistent flux, peak flux of bursts and their errors are obtained by fitting models with {\it Xspec} command "cflux".
}
\begin{minipage}{14cm}
\begin{tabular}{lccc|lccc}
\hline
\hline

Observation ID & Modified Julian Day    &   Persistent Flux            &     Peak Flux           &   Observation ID & Modified Julian Day    &   Persistent Flux            &     Peak Flux                       \\
               &                        &  ($\rm {10}^{-9}\,erg\ s^{-1}\ {cm}^{-2}$)   &   ($\rm {10}^{-9}\,erg\ s^{-1}\ {cm}^{-2}$)        &                  &                        &   ($\rm {10}^{-9}\,erg\ s^{-1}\ {cm}^{-2}$)   &    ($\rm {10}^{-9}\,erg\ s^{-1}\ {cm}^{-2}$)                      \\

\hline
   30054-04-02-000 & 50971.70  &  0.996 $\pm $  0.010 & 28.8  $\pm $ 0.9 &  90043-01-01-01 &50971.70 &  1.507 $\pm $  0.005 & 25.5  $\pm $ 0.8 \\
    30054-04-02-01 & 50971.23  &  1.012 $\pm $  0.004 & 27.7  $\pm $ 0.9 &  90043-01-01-02 &50971.23 &  1.581 $\pm $  0.007 & 26.5  $\pm $ 0.8 \\
    30054-04-02-02 & 50972.18  &  1.126 $\pm $  0.007 & 28.1  $\pm $ 0.9 & 90043-01-01-020 &50972.18 &  1.577 $\pm $  0.013 & 26.0  $\pm $ 0.8 \\
    30054-04-03-02 & 50976.42  &  1.014 $\pm $  0.004 & 27.5  $\pm $ 0.7 &  90043-01-02-00 &50976.42 &  1.617 $\pm $  0.009 & 24.7  $\pm $ 0.9 \\
    30060-03-01-01 & 50988.83  &  0.991 $\pm $  0.006 & 27.1  $\pm $ 0.7 & 90043-01-02-001 &50988.83 &  1.578 $\pm $  0.026 & 25.6  $\pm $ 0.9 \\
    50035-01-01-01 & 51724.88  &  1.309 $\pm $  0.007 & 24.6  $\pm $ 0.7 & 90043-01-02-001 &51724.88 &  1.578 $\pm $  0.026 & 24.2  $\pm $ 0.8 \\
    50035-01-02-00 & 51725.71  &  1.176 $\pm $  0.019 & 27.3  $\pm $ 0.7 &  90043-01-02-01 &51725.71 &  1.636 $\pm $  0.007 & 24.7  $\pm $ 0.8 \\
    50035-01-02-00 & 51725.88  &  1.176 $\pm $  0.019 & 26.8  $\pm $ 0.7 &  90043-01-02-01 &51725.88 &  1.636 $\pm $  0.007 & 25.1  $\pm $ 0.8 \\
    50035-01-02-02 & 51726.72  &  1.364 $\pm $  0.009 & 26.0  $\pm $ 0.7 & 90043-01-03-000 &51726.72 &  1.620 $\pm $  0.021 & 26.5  $\pm $ 0.8 \\
    50035-01-02-04 & 51728.77  &  1.359 $\pm $  0.014 & 26.0  $\pm $ 0.7 &  90043-01-03-01 &51728.77 &  1.551 $\pm $  0.011 & 25.7  $\pm $ 0.8 \\
    50035-01-03-00 & 51811.75  &  1.324 $\pm $  0.016 & 26.5  $\pm $ 0.8 &  90043-01-03-01 &51811.75 &  1.551 $\pm $  0.011 & 24.7  $\pm $ 0.8 \\
    50035-01-03-08 & 51813.49  &  1.361 $\pm $  0.012 & 24.9  $\pm $ 0.7 & 91017-01-01-000 &51813.49 &  1.537 $\pm $  0.028 & 23.5  $\pm $ 0.8 \\
    50035-01-03-09 & 51814.01  &  1.417 $\pm $  0.017 & 26.4  $\pm $ 0.8 &  91017-01-01-05 &51814.01 &  1.600 $\pm $  0.015 & 25.7  $\pm $ 0.8 \\
    50035-01-03-10 & 51814.35  &  1.397 $\pm $  0.007 & 26.0  $\pm $ 0.8 &  91017-01-02-01 &51814.35 &  1.580 $\pm $  0.011 & 23.8  $\pm $ 0.8 \\
   50035-01-03-120 & 51813.15  &  1.368 $\pm $  0.011 & 26.1  $\pm $ 0.7 & 91017-01-02-010 &51813.15 &  1.543 $\pm $  0.006 & 25.9  $\pm $ 0.8 \\
    70044-01-01-00 & 52484.57  &  1.578 $\pm $  0.017 & 27.6  $\pm $ 0.8 & 91017-01-02-010 &52484.57 &  1.543 $\pm $  0.006 & 24.8  $\pm $ 0.8 \\
   70044-01-01-000 & 52484.42  &  1.717 $\pm $  0.038 & 24.7  $\pm $ 0.8 &  91017-01-02-03 &52484.42 &  1.587 $\pm $  0.009 & 26.6  $\pm $ 0.8 \\
    70044-01-01-02 & 52485.01  &  1.615 $\pm $  0.024 & 26.4  $\pm $ 0.7 &  91017-01-02-09 &52485.01 &  1.583 $\pm $  0.015 & 25.7  $\pm $ 0.8 \\
    70044-01-02-00 & 53205.20  &  1.580 $\pm $  0.013 & 24.3  $\pm $ 0.8 &  91017-01-02-11 &53205.20 &  1.557 $\pm $  0.045 & 25.2  $\pm $ 0.8 \\
    70044-01-02-01 & 53206.06  &  1.514 $\pm $  0.012 & 26.3  $\pm $ 0.8 & 92031-01-01-000 &53206.06 &  1.257 $\pm $  0.008 & 20.2  $\pm $ 0.6 \\
    70044-01-04-00 & 53966.37  &  1.654 $\pm $  0.010 & 24.5  $\pm $ 0.8 & 92031-01-01-000 &53966.37 &  1.257 $\pm $  0.008 & 20.8  $\pm $ 0.7 \\
    80048-01-01-00 & 52736.53  &  1.417 $\pm $  0.009 & 23.6  $\pm $ 0.7 & 92031-01-01-001 &52736.53 &  1.227 $\pm $  0.014 & 17.8  $\pm $ 0.6 \\
   80048-01-01-010 & 52736.67  &  1.410 $\pm $  0.006 & 24.4  $\pm $ 0.8 & 92031-01-01-010 &52736.67 &  1.251 $\pm $  0.005 & 17.9  $\pm $ 0.5 \\
   80048-01-01-010 & 52736.80  &  1.410 $\pm $  0.006 & 24.0  $\pm $ 0.6 & 92031-01-01-010 &52736.80 &  1.251 $\pm $  0.005 & 18.7  $\pm $ 0.5 \\
    80048-01-01-04 & 52738.48  &  1.348 $\pm $  0.007 & 24.5  $\pm $ 0.7 & 92031-01-01-011 &52738.48 &  1.283 $\pm $  0.024 & 20.5  $\pm $ 0.7 \\
    80048-01-01-07 & 52738.61  &  1.397 $\pm $  0.014 & 23.6  $\pm $ 0.7 & 92031-01-01-011 &52738.61 &  1.283 $\pm $  0.024 & 19.3  $\pm $ 0.7 \\
    80048-01-01-07 & 52738.75  &  1.397 $\pm $  0.014 & 32.4  $\pm $ 0.9 & 92031-01-01-011 &52738.75 &  1.283 $\pm $  0.024 & 20.2  $\pm $ 0.7 \\
    80049-01-01-00 & 52820.56  &  1.507 $\pm $  0.009 & 24.1  $\pm $ 0.5 &  92031-01-02-00 &52820.56 &  1.282 $\pm $  0.013 & 20.2  $\pm $ 0.6 \\
    80049-01-02-00 & 52944.40  &  1.425 $\pm $  0.010 & 26.4  $\pm $ 0.8 & 92031-01-02-000 &52944.40 &  1.231 $\pm $  0.007 & 19.8  $\pm $ 0.7 \\
    80049-01-03-02 & 52834.39  &  1.515 $\pm $  0.010 & 25.1  $\pm $ 0.8 & 92031-01-02-000 &52834.39 &  1.231 $\pm $  0.007 & 19.6  $\pm $ 0.6 \\
    80049-01-03-03 & 52835.31  &  1.523 $\pm $  0.013 & 26.9  $\pm $ 0.7 &  92703-01-04-02 &52835.31 &  1.561 $\pm $  0.009 & 24.4  $\pm $ 0.8 \\

\hline
\end{tabular}
\label{table1}
\end{minipage}
\end{table*}

\end{document}